\documentclass[a4paper,12pt]{article}
\usepackage[all]{xy}
\usepackage{amsfonts}
\usepackage{amssymb}
\usepackage{epsfig}
\usepackage{graphicx}
\usepackage{amsmath}

\def\beq{\begin{equation}}
\def\eeq{\end{equation}}
\def\bea{\begin{eqnarray}}
\def\eea{\end{eqnarray}}

\catcode`\@=11
\def\numberbysection{\@addtoreset{equation}{section}
        \def\theequation{\thesection.\arabic{equation}}}

\numberbysection

\begin{document}

\begin{titlepage}
\begin{center}
\vskip .5 in {\Large\bf  Topological phase transition in a RNA
model}

{\Large\bf in the de Gennes regime } \vskip 0.3in

Mat\'{\i}as G. dell'Erba\\

{\em Instituto de F\'{\i}sica de Mar del Plata, IFIMAR (CONICET-UNMdP)}
\\
and
\\
{\em Departamento de F\'{\i}sica, Facultad de Ciencias Exactas y
Naturales, Universidad Nacional de Mar del Plata, Funes 3350,
(7600), Mar del Plata, Argentina}
\\
\vskip 0.3in Guillermo~R.~Zemba\footnote{ Member of CONICET,
Argentina.}
\\
{\em Facultad de Ciencias Fisicomatem\'aticas e Ingenier\'{\i}a,
Pontificia Universidad Cat\'olica Argentina, Av. A. Moreau de Justo 1500,
Buenos Aires, Argentina}
\\
and
\\
{\em Departamento de F{\i}sica, C.N.E.A.
Av.Libertador 8250, (1429) Buenos Aires, Argentina}
\end{center}
\vskip .5 in

\begin{abstract}
\noindent
We study a simplified model
of the RNA molecule proposed by
G. Vernizzi, H. Orland and A. Zee in the
regime of strong concentration of positive
ions in solution. The model considers a
flexible chain of equal bases that can pairwise interact
with any other one along the chain,
while preserving the property of saturation of the
interactions.
In the regime considered, we observe the emergence of a
critical temperature $T_c$ separating two
phases that can be characterized by
the topology of the predominant configurations:
in the large temperature regime,
the dominant configurations of the molecule
have very large genera (of the order of the
size of the molecule), corresponding to a
complex topology, whereas in the opposite
regime of low temperatures,
the dominant configurations are simple and
have the topology of a sphere.
We determine that this topological phase transition
is of first order and provide an
analytic expression for $T_c$.
The regime studied for this model exhibits analogies
with that for the dense polymer systems studied by de Gennes.
\end{abstract}


PACS numbers: 87.14gn , 02.10.Yn , 87.15.Cc

\vfill
\end{titlepage}
\pagenumbering{arabic}


\section{Introduction}

Recent advances in mathematical models of RNA and DNA molecules
have led to stimulating new studies of their complex spatial
properties and their relation with the environmental variables (
see, {\it e.g.}
\cite{Tinoco,Zee2,Muller,Hyeon,Pande} ). In
this paper we consider the exactly solved model for a RNA molecule
presented in \cite{Zee1}, which consists of a homopolymer of length
$L$, with an infinitely flexible backbone, and in which any
arbitrary pair of bases is allowed to (pairwise) interact. This
combinatorial model preserves the important property of {\it saturation}
of the interactions \cite{Zee2,Vernizzi} of the actual RNA
molecule, but does not include both the geometric and energetic
aspects of the real molecules. Due to its simplicity, the model
is exactly solved, {\it i.e.}, it allows for analytical
expressions of statistical and topological interesting quantities,
such as the partition function and its topological expansion
\cite{Nuestro}. A crucial ingredient of the model is the
introduction of an extra
degree of freedom, $N$, such that a random $N
\times N$ hermitian matrix is added to each base position along 
the chain. A physical
interpretation can be given to this parameter {\it a posteriori}
\cite{Zee1,Nuestro,Zee5}: $1/N$ plays a role in the model that
can be associated to the concentration
of positive ions (like $\rm{Mg}^{++}$) in solution for real
molecules. These positive ions are responsible for
the overall electric charge neutralization of the system
compensating the negative charge in the
phosphate groups in the RNA molecule. The neutralization of the
phosphate ions provide the physical mechanism for the folding of the
molecule \cite{Lindahl,RNAworld,Misra,Draper,Travers,Hackl}.
Moreover, the algebraic power of $1/N^2$ in the topological 
expansion of the
partition function of the model is the genus of the diagrams
associated to the different configurations \cite{Zee1,Nuestro,'t
Hooft}. 
These diagrams encode the real spatial structure of the
molecule, in the sense that they are analogous to the
secondary (or planar) structure of the molecule, from which the
tertiary (or spatial) one is obtained \cite{Zee1}. 
Therefore, the parameter $N$ works as a regulator
that controls the topology of the molecule within the model.
To the best of our
knowledge, the regime of $N \to 0 $ has not been received much
attention in the theoretical models considered in the literature,
and in this paper we address some questions arising in this regime.
We identify this regime with the dense polymer phase of the $O(n)$ vector
model of de Gennes \cite{de Gennes},
which is described by the limit $n\to 0$.
Furthermore, we are not aware of
any experimental works in real RNA molecules in the
corresponding laboratory regime.

In a previous work \cite{Nuestro}, we have provided an exact
analytic expression for the partition function of the model, which
has been used to study its thermodynamic properties in the regime
$N \geq 1 $, {\it i.e.}, that of small ion concentration in the
surrounding environment. However, some of the expressions that we have 
obtained
in that paper are analytical in $N$ and are, therefore, useful
to explore other regimes than the one we have chosen to discuss
in that work. In this paper we study (for a wide range of
temperatures) the complementary
regime $N \to 0 $ , which can be characterized as one with large
positive ion concentration in the surrounding medium. 
We observe the emergence of a critical temperature
$T_c$ separating two regimes with different spatial
configurations: in the large temperature regime,
the dominant configurations of the molecule
have very large genera (of the order of the
size of the molecule), corresponding to a
complex topology, whereas in the opposite
regime of low temperatures,
the dominant configurations are simple and
have the topology of a sphere. 
This transition is not related {\it a-priori}
to the so-called
{\it coil-globule transition} in which
the molecule adopts the
shape of an elongated coil for low temperatures, whereas it 
assumes a compact globule shape in the large temperature
regime. This latter transition is
considered to be of first order \cite{Muller, Zee4, Noguchi, Ueda,
Doniach}.

Some recent work in the literature has recently addressed
related issues. In \cite{Zee4},
a study of the distribution of genera of
pseudoknotted configurations in the $T$-dependent phase transition
for a self-avoiding homopolymer on a lattice has been considered.
A study of the behavior of the distribution of genera for the model
introduced in \cite{Zee1} with external perturbations
has been the subject of Refs. \cite{Garg1,Garg2}.
Moreover, in \cite{Garg1}, a `structural transition' dependent on the external
perturbation has been also discussed.

This paper is organized as follows: in Section 2 we
reconsider the model presented in
\cite{Zee1} and recall some exact results, such as the partition function and
its topological expansion \cite{Nuestro}.
Section 3 is devoted to the study of 
the regime $N\to 0$ and the topological phase
transition: we first consider the mathematical
setting and approximations that are needed to analytically
treat the phase transition.
We then apply this preliminary study to the partition
function of the model to make explicit the emergence of the phase transition
and we give an analytic expression for the critical temperature.
In section 4, we discuss the
thermodynamic properties of the system in the limit $N \to 0$. We
calculate the free energy per particle, the entropy and the latent
heat. From the last thermodynamic quantity, we show that the phase transition is
of first order. Finally, we give our conclusions.


\section{Topological expansion of the partition function}

We start by reviewing some of the properties of the model proposed
by G. Vernizzi, H. Orland and A. Zee \cite{Zee1}. The model
considers a chain of bases of one type only, such that the
interaction energy between any pair of bases is a constant
$\epsilon$. A given base can interact with any other base in the
chain, but preserving the {\it saturation property}, which
excludes interactions among three or more bases. We consider that
this property is one of the essential ingredients of the
simplified model. Therefore, all the Boltzmann factors
$v=\exp{(-\epsilon/\kappa T)}$ (where $T$ is the absolute
temperature and $\kappa$ is the Boltzmann constant) are equal as
well. At each base site, a random $N \times N$ hermitian matrix
is added as the relevant degree of freedom. We consider
this feature to be a second essential ingredient of the model. The
configurational partition function $Z$ of a chain of length $L$ is: 
\bea
Z(L,N,T)&=&\frac{1}{N}\langle{\rm
tr}(1+\frac{1}{L^{1/2}}\varphi)^{L}\rangle\\
&=&\int d\varphi\
e^{-\frac{N}{2Lv}{\rm tr}\varphi^{2}}\frac{1}{N}{\rm
tr}(1+\frac{1}{L^{1/2}}\varphi)^{L} {\large /}
{\int d\varphi\ e^{-\frac{N}{2Lv}{\rm tr}\varphi^{2}}}, \label{Zvalormedio}
 \eea
 \noindent
where $\varphi $ represents a collective degree of freedom (a sort of
center-of-mass random matrix). The simple form of
(\ref{Zvalormedio}) is a consequence of the symmetry of the matrix
potential that reduces the original integration over $L$ matrices
to one integration over $\varphi$ \cite{Zee1}. Applying standard results of
random matrices to (\ref{Zvalormedio}) one obtains:
 \beq
 Z(L,N,T)=\sum_{k=0}^{[L/2]} d_{k}(L,N)\ e^{-\epsilon_{k}/\kappa T}\ ,
 \label{PartFunctTrad}
 \eeq
\noindent
where the symbol $[L/2]$ means the integer part of $L/2$,
$\epsilon_{k}= k \epsilon$ and
 \beq
d_{k}(L,N)=\sum_{j=0}^{k} \left(%
\begin{array}{c}
  L \\
  2k \\
\end{array}%
\right) \left(%
\begin{array}{c}
  k \\
  j \\
\end{array}%
\right) \left(%
\begin{array}{c}
  N \\
  j+1 \\
\end{array}%
\right)\frac{(2k)!}{2^{k-j} \ k! \ N^{k+1}}\ .
 \label{Deg}
 \eeq
\noindent
From (\ref{PartFunctTrad}) we may compute $Z$
exactly (for each $L$), as a function of $N$ and $T$. The spectrum of
the system has $[L/2]+1$ energy levels, with energies: $0, \epsilon,
2\epsilon, \ldots ,[L/2] \epsilon$ and the degeneracy of the
$k-th$ level is $d_{k}(L,1)$.

As we have mentioned before, the power of $1/N^{2}$ yields the genus
$g$ of the diagram obtained from the chain by joining all interacting bases
by a `photonic line' \cite{Zee1}, that is, the minimum number of handles of the
surface on which the diagram can be drawn without crossings. In
\cite{Nuestro}, we have written the partition function of the model in the form
of a \emph{topological expansion} \cite{'t Hooft,Zee1,Zee2},{\it i.e.},
as a power series in $1/N^2$, where the coefficients take
into account all the Feynman diagrams with the same topological
character:
 \beq
 Z(L,N,T)=\sum_{g=0}^{\infty} z_{g}(L,T) \frac{1}{N^{2g}}\ .
 \label{ExpTop}
 \eeq
\noindent
Here $z_{g}(L,\infty)$ is the number of planar diagrams
that can be drawn on a topological surface of genus $g$ for a
molecule of size $L$. Note that, as a function of $T$,
$z_{g}(L,T)$ is the partition function of the system living on the
topological surface of genus $g$. The coefficients $z_{g}(L,T)$
are given by:
 \beq
z_{g}(L,T)=\sum_{k=0}^{[L/2]}\sum_{j=k-2g}^{k}
 \frac{L! \ 2^{j-k} \ S_{j+1}^{(k+1-2g)}}{(L-2k)! \ (k-j)! \ j! \
 (j+1)!} \
e^{- \epsilon_{k}/\kappa T}\ , \label{ZxTop}
 \eeq \noindent
where $S_{j}^{(m)}$ is the Stirling number of the first kind
\cite{Gradshteyn,Stirling} with parameters $m,j$ ($S_{j}^{(m)}=0$
if $m>j$ or if $j\leq0$). In the limit $T\rightarrow\infty$,
$z_{g}(L,T)$ coincides with the coefficients $a_{L,g}$ of Ref.
\cite{Zee1}. Using the property of the Stirling numbers mentioned
in this paragraph, we see from (\ref{ZxTop}) that the
maximum genus of a diagram for a given $L$ is $[L/4]$ and,
therefore, $g\leq[L/4]$.


\section{The partition function in the limit $N \to 0$}

\subsection{Preliminary study of the analytic behavior of the partition
function} \label{preliminar}

In a previous paper, we have considered the limit $N \to \infty$
of the partition function (\ref{PartFunctTrad}) \cite{Nuestro},
for which we gave explicit expressions. As
a consequence, we have verified the consistency of interpreting the
parameter $1/N$ as being proportional to the density of positive
ions in the media surrounding the molecule, as has been suggested
in \cite{Zee1}. Therefore, the range of the analysis considered in
\cite{Nuestro} applies to media with small concentration of
positive ions. It is natural to consider as well the opposite
situation of large positive ion concentration in the surrounding
medium, which corresponds to the limit $N \to 0 $.
We are going to
show later on that this regime is characterized by the existence
of a critical temperature and a phase transition.

\begin{figure}[h]
\epsfysize=6cm \centerline{\epsfbox{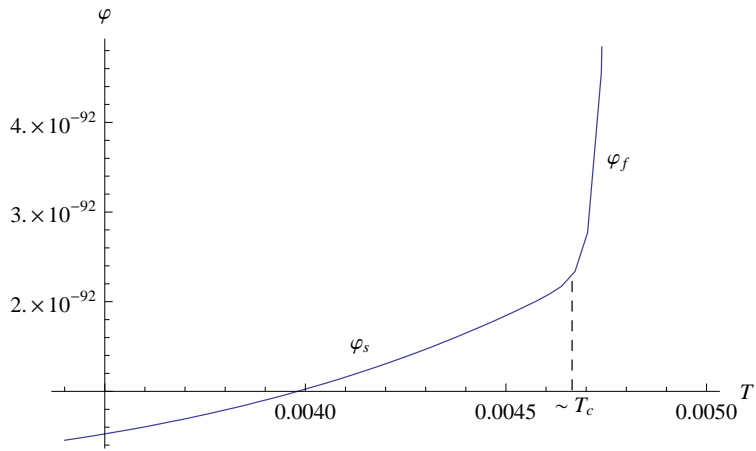}}
\caption{Plot of $\phi(N,T)=e^{-1/T}+ N T^5$ around of $T_c \simeq
0.00466$ with $N=10^{-80}$.The plot has been done numerically.} \label{TransFase}
\end{figure}

From the mathematical point of view, studying this limit requires
a careful handling of the partition function considered as
an analytic function of several variables. Here we discuss a mathematical
scheme that allows us to obtain simple analytic expressions in this
limit. We first consider a general function $\phi$ of two variables $N$
and $T$ (which will be ultimately related to the free energy)
which admits a decomposition of the form:
\begin{equation}
\phi(N,T) = \phi_{f}(T)+\phi_{s}(N,T)\ ,
\end{equation}
where $\phi_{f}$ ($f$ means `fast') indicates a function whose
asymptotic grow in the variable $T$ is in the class of the
exponential function, whereas the function $\phi_{s}$ ($s$ means
`slow') denotes a function that grows slower than the class of the
exponential function ({\it e.g.}, polynomial, logarithmic
classes). Depending on the value of $N$ (which we consider fixed
at $N_0$) there exists a small range of values of $T$ for which
$O(\phi_{f}(T)) = O(\phi_{s}(N_0,T))$ for values of $T$ nearby a
critical value $T_c$. At this point we can anticipate the idea of
defining the critical parameter $T_c$ as the value of $T$ for
which both the slow and fast parts of the function
$\phi$ become equal. We shall further develop on this idea below.
The parameter $T_c$ emerges naturally in the case
when the function $\phi$ is the free energy
where it plays the role of a critical temperature. Given the asymptotic
behavior of $\phi_{f}$, as soon as $T$ crosses $T_c$, the order of
$\phi_{f}$ varies rapidly, increasing or decreasing, depending of
the sense in which $T$ crosses $T_c$, so that (for example):
\bea
    \phi(N_0,T>T_c) &\simeq& \phi_{f}(T)\ ,\\
    \phi(N_0,T<T_c) &\simeq& \phi_{s}(N_0,T)\ .
 \eea
Therefore, $\phi(N_0,T)$ has different analytical behavior
according to the region in which $T$ is located with respect to
$T_c$. The change in $\phi$ between these two regions becomes more
pronounced when $\phi_{f}$ varies more rapidly. In Fig.
\ref{TransFase} we show an example of this behavior for the 
cases $\phi_f=e^{-1/T}$ and $\phi_s=N T^5$ (in this example $T$ and $N$ are two
dimensionless variables) , in the vicinity of
$T_c \simeq 0.00466$ (the plot has been done numerically):
for $T > T_c$, $\phi$ is dominated by $\phi_{f}$ and for $T
< T_c$ it is dominated by $\phi_{s}$.

If we assume that $\phi_{f}$ is of the form $\phi_{f}(T) \sim
e^{-1/T}$, the change is even more noticeable for small values of $T$ . If we
had $T_c \to 0$, the derivative $\phi'(N_0,T = T_c)$ would not
exist. We shall correlate the absence of the derivative of $\phi$ at
the critical point with the appearance of a phase transition of
the system for $T_c = 0$.

\subsection{The critical temperature and the phase transition}

We now consider the topological expansion in $1/N^2$
(\ref{ExpTop}) of the partition function $Z$ of the system.
Considering that the maximum genus of the diagrams of all possible
configurations is $[L/4]$ (as we have mentioned above), we
rewrite $Z$ as a polynomial in the variable $N^2$ times a
$N$-dependent factor
which is divergent in the $N \rightarrow 0$ limit:
\bea
Z(L,N,T) &=&\frac{1}{N^{2[L/4]}} Z_R(L,N,T)\ ,\\
Z_{R}(L,N,T) &=& z_{[L/4]}+z_{[L/4]-1}N^2+\cdots +z_{0}N^{2[L/4]}\ . \label{ZR}
 \eea
Here the subindex $R$ indicates that the function is {\it regular}
in the limit $N \rightarrow 0$. The coefficients $z_{g}$ were
given in the previous section \cite{Nuestro}, and can be written as:
\begin{equation}
    z_{g}(L,T)=\sum_{k=0}^{[L/2]}
    r_{k,g}(L) \hspace{0.1cm} e^{-k \epsilon/ \kappa T} , \label{CoefExpTop}
\end{equation}
where the coefficients $r_{k,g}(L)$ are given by
 \begin{equation}
    r_{k,g}(L)=\frac{L!}{(L-2k)! 2^k} \sum_{j=k-2g}^{k} \frac{2^{j} S_{j+1}^{(k-2g+1)}}{(k-j)! j! (j+1)!}\ , \label{rkg}
 \end{equation}
and $S_{i}^{(j)}$ are the Stirling numbers of the first kind with
parameters $i$ and $j$. Using the property
$S_{i}^{(0)}=\delta_{i0}$ \cite{Abramowitz}, we rewrite the
partition function on the topological surface of genus $[L/4]-n$
as:
 \begin{equation}
z_{[L/4]-n}(L,T)=\sum_{k=[L/2]-2n+\delta}^{[L/2]}
    r_{k,[L/4]-n}(L) \hspace{0.1cm} e^{-k \epsilon /\kappa T}\ ,
 \end{equation}
where $\delta=-1$ for $[L/2]$ odd and zero otherwise. Without loss
of generality, we shall consider $[L/2]$ even from now on.
Substituting the coefficients $z_{g}$ in (\ref{ZR}), we obtain the
regular part of the partition function (\textit{RZ}):
 \begin{multline}
Z_{R}(L,N,T)= e^{-[L/2] \epsilon /\kappa T} \{
r_{[L/2],[L/4]} + (r_{[L/2]-2,[L/4]-1}e^{2 \epsilon/ \kappa
T} + r_{[L/2]-1,[L/4]-1}e^{\epsilon/ \kappa T} \\+
r_{[L/2],[L/4]-1}) N^2 + \cdots + (r_{0,0}e^{[L/2]
\epsilon/ \kappa T} \\ + r_{1,0}e^{([L/2]-1)
\epsilon /\kappa T} + \cdots + r_{[L/2],0} )
N^{2[L/4]}\}\ .    \label{Zcompleta}
 \end{multline}
Note that $r_{0,0}=1$ since it counts the number of
configurations with lowest energy and genus, that is, the
unique configuration without interaction among bases.
In order to treat (\ref{Zcompleta}) analytically in
what follows, we shall consider
two approximations on it, valid in the limit $N \ll 1$.
We have studied the analytic behavior of (\ref{Zcompleta})
using the program $Mathematica$
\cite{Mathematica}, within the range $N \lessapprox
10^{-20}$ in the concrete examples considered.
For these examples,
the range of sizes of the system is $L \sim 10^{2}-10^{4}$ and
that for the temperature is $T \sim 10^{-4}-10^{-2}$.

First, we observe that for the ranges of $L$ and $T$ considered,
the terms proportional to $N^{2g}$ are dominated by exponential
factors ({\it e.g.}, $r_{k,g}/r_{k-1,g} \lessapprox 10^7$ and
$e^{\frac{\epsilon}{\kappa T}} \gtrapprox 10^{43}$ ($\epsilon =
\kappa = 1$)). For large values of $T$ the most important term in
the sum (\ref{Zcompleta}) is the first one (up to order zero in
$N$), whereas for small values of $T$, the dominant term in the sum
is the one proportional to $N^{2[L/4]}$ ($= N^{[L/2]}$, for
$[L/2]$ even). To keep analytic expressions as simple as possible,
we retain only the most important term proportional to $N^{[L/2]}$
in (\ref{Zcompleta}), {\it i.e.}, $e^{[L/2] \epsilon / \kappa
T}$ and write:
 \begin{equation}
Z_{R}(L,N,T) \simeq  r_{[L/2],[L/4]} e^{-[L/2]
\epsilon / \kappa T} + N^{[L/2]}\ . \label{ZTA}
 \end{equation}
We refer to (\ref{ZTA}) as the Thermal Approximation (\textit{TA})
for the partition function, since the relevance of the terms in
$Z_R$ vary significatively when $T$ crosses $T_c$, that is, when
$T$ satisfies $O(r_{[L/2],[L/4]} e^{-[L/2] \epsilon / \kappa
T})=O(N^{[L/2]})$ (see \ref{preliminar}).

\begin{figure}[h]
\epsfysize=7cm \centerline{\epsfbox{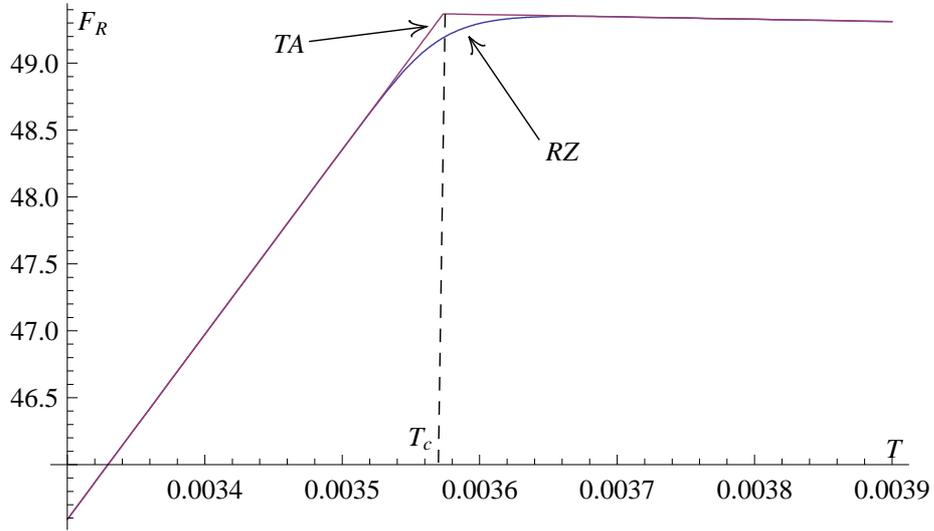}}
\caption{Typical behavior of $F_{R}$ calculated from \textit{RZ}
and \textit{TA} as a function of $T$ for $L=100$ and
$N=10^{-120}$.} \label{F2F1}
\end{figure}

Next, we specify
the definition of $T_c$ with more detail. It is clear that
(\ref{ZTA}) is a broader approximation to (\ref{Zcompleta}), since
it excludes many terms that are relatively important when the two
terms in (\ref{ZTA}) are of the same order, {\it i.e.}, when $T$
is closer to $T_c$. However, we will see later how to minimize
the difference between (\ref{ZTA}) and (\ref{Zcompleta}), showing
the conditions under which \textit{RZ} becomes close to \textit{TA}.
From the \textit{TA} we calculate the regular part of free energy
from $F = - \kappa T \ln Z$:
\begin{equation}
F_{R}(L,N,T) \simeq - \kappa T \ln\{ r_{[L/2],[L/4]}
e^{-[L/2] \epsilon /\kappa T} + N^{[L/2]}\}\ .
\label{F2aprox}
\end{equation}
In Fig. \ref{F2F1} we show a plot of the thermal approximation
for the free energy \textit{TA} against the temperature. Two
regimes can be clearly differentiated, separated by $T=T_{c}$: for
$T<T_{c}$, the second term in the logarithm function of
(\ref{F2aprox}) is much larger than the first one, and can be
therefore neglected, with opposite behavior for $T>T_{c}$. These
plots show two interesting characteristics: (\ref{ZTA}) behaves
linearly below and above $T_{c}$ and the slope for $T<T_{c}$ is
higher than that for the case with $T>T_{c}$. We therefore
identify a phase transition occurring for $T=T_{c}$, which we
identify as the critical temperature. As it is customary for a finite
system such as the model for the RNA molecule studied here, the term
`phase transition' should be interpreted as a `strongly
cooperative phenomenon' \cite{FenomenoCooperativo}.

In a real RNA molecule,
the large negative electric charge of the phosphate ions,
prevents it to fold onto a compact structure.
The concentration of positive ions in solution, such as
$\rm{Mg^{++}}$, neutralize the phosphate ions making possible the
folding of the RNA molecule \cite{Mg++,Draper,Lindahl}.
In the context of the model studied in this paper, if we
interpret $1/N$ as the concentration of $\rm{Mg^{++}}$ in solution
\cite{Nuestro,Zee2,Zee5}, we see from (\ref{F2aprox}) that
increasing this concentration favors the formation of RNA
structures with large genus, and viceversa.

The asymptotic straight lines in Fig. \ref{F2F1} are obtained by
retaining only the leading term in the logarithm of (\ref{F2aprox}) when
varying the temperature around $T_{c}$. Therefore, the analytic
expressions for these asymptotic straight lines are given by:
 \bea
F_R\left(L,N,T<T_c\right) &\simeq& - \kappa [L/2] \ln \{ N\} T
\label{F-Tc}\ , \\
F_R\left(L,N,T>T_c\right) &\simeq& -\kappa  \ln \{r_{[L/2],[L/4]}\}T
+ [L/2]\epsilon\ , \label{F+Tc}
 \eea
and we obtain an analytic expression for the critical
temperature equating the two
terms in the logarithm of (\ref{F2aprox}):
\begin{equation}
T_c(L,N)=\frac{\epsilon }{\kappa \ln \left\{
(r_{[L/2],[L/4]})^{\frac{1}{[L/2]}}\frac{1}{N}\right\}} \label{Tc}
 \end{equation}
As we have mentioned in Sec. \ref{preliminar} and show in Fig.
\ref{F2F1}, the regular free energy $F_R$ undergoes a change in
its analytic behavior when the temperature reaches the critical
value $T_c$. Below $T_c$, $F_R$ is dominated by the slowly varying function
($N^{[L/2]}$), whereas above $T_c$, the rapidly varying function ($
r_{[L/2],[L/4]} e^{-\frac{[L/2] \epsilon}{\kappa T}}$) dominates.
The change in the behavior of $F_R$ is sharp in the \textit{TA},
while for \textit{RZ} it is obtained for a small range of
temperatures because \textit{RZ} presents additional terms
that acquire relative importance when $T$ is close to $T_c$.

It can be seen numerically from (\ref{rkg}) that the function of $L$ in $T_c$
behaves linearly as $(r_{[L/2],[L/4]})^{\frac{1}{[L/2]}} = 
L/e - 4.5521$. For large $L$ ($L > 200$), we can write the critical
temperature as:
 \begin{equation}
T_c(L,N) \simeq \frac{\epsilon }{\kappa \ln \left ( L /eN \right )}\ . \label{TcAprox}
\end{equation}
Note that (\ref{TcAprox}) possess the symmetry $T_c(L,N)=T_c(a L,a
 N)$ for $a>0$ a real parameter, that is, $T_c$ is scale invariant.
Therefore, in the large $L$ limit, we have that $T_c(L,N) = T_c(L/N)$
Furthermore, there exists a natural cut-off for $N$ such that $T_c \geq 0$:
\begin{equation}
N < (r_{[L/2],[L/4]})^{\frac{1}{[L/2]}} \simeq  L/e\ .
\end{equation}
This is a consequence of (\ref{Tc})
and it is consistent with the condition stating that 
the slope of the asymptotic
straight line for $T < T_c$ must be larger than
that for $T > T_c$.
However, this cut-off is not very restrictive because we are
interested in the the limit $N \ll 1$ and $L$ is at least of $O(1)$ .

Moreover, there exist further and more restrictive conditions on
$N$, implied by both the  thermodynamic and large size limits.
Given that \textit{RZ} should approach \textit{TA} for large $L$,
a relationship between $N$ and $L$ should exist. When $L$ increases
and $N$ is fixed, the difference between the analytic  expressions
for \textit{RZ} and \textit{TA} also increases. In order to obtain
a convergence between the two in the thermodynamic limit, we make
$N$ a $L$-dependent variable as follows: for each $L$, there
exists an upper limit for $N$ below which the phase transition
exists. This can be seen by demanding that the scenario discussed
in the previous section actually occurs. For real molecules, this
would mean that there should exist a minimal ($L$-dependent)
concentration of $\rm{Mg}^{++}$ for which the phase transition
exists. As a consequence of this restriction, in the thermodynamic
limit $L\to\infty$ then  $N\to 0$ , implying $T_c \rightarrow 0$.
In this limit, the derivative of $F_R$ at $T = T_c$ (see
(\ref{F2aprox})) does not exist and there is a {\it bona fide}
phase transition.

Furthermore, the limit $N \rightarrow 0$ in the model we have
considered can be associated with
the dense polymer phase of the $n$ vector model
which arises for $n\to 0$ as has been discussed by
de Gennes \cite{de Gennes}.
The correspondence is established between the degrees
of freedom of the $O(n)$ vector model, which are
$n$-component spin vectors, and the degrees of
freedom of the RNA model of \cite{Zee1}, given by
by $N \times N$ hermitian matrices.
The limit $n \rightarrow
0$ in the de Gennes $O(n)$ model corresponds to a
high density polymer phase, which is naturally
associated to the high concentration of $\rm{Mg}^{++}$
(or other positive ions such as $\rm{K}^+$)
in solution phase of the RNA model.

\begin{figure}[h]
\epsfysize=3.5cm \centerline{\epsfbox{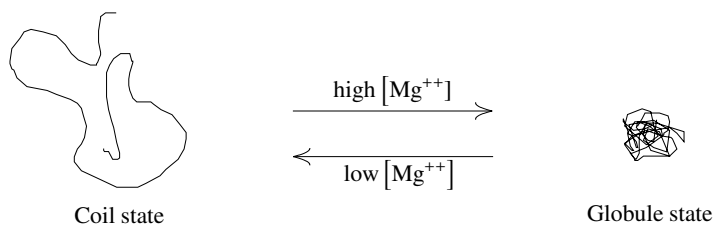}}
\caption{Phase transition between coil and globule states when $T$
crosses $T_c$. The arrows indicate the direction in which $T$ increases,
to right (left) for high (low) concentration of Mg$^{++}$. This is not to
confused with the topological phase transition, as explained in the text.}
\label{CoilvsGlobule}
\end{figure}

From the previous discussion of the RNA model, the $T$-dependent
phase transition involves a topology change in the spatial
configurations of the molecule, which goes from one with
genus zero for $T < T_c$ to another one with large genus
$[L/4]$ for $T > T_c$. The difference in the topology of the
configurations of the molecule when $T$ crosses $T_c$ justifies the
use of the term `topological' for describing the nature of the phase
transition. Note that this transition does not correspond to the
coil-globule transition in polymers studied in \cite{Zee4}, in
which the genus of the configurations decreases with increasing
temperature (see Fig. \ref{CoilvsGlobule}).

This is due to the fact that both transitions take place in
different regimes: whereas in \cite{Zee4} the RNA molecule is
surrounded by a dilute solution of $\rm{Mg}^{++}$, for the case
we have discussed the molecule is imbedded in a medium
with large concentration of positive ions.

\section{Thermodynamic properties in the $N \to 0$ limit}

From the analytic results for the partition function obtained in
the previous section, we now calculate some thermodynamic
quantities that we will use to better describe the configurations
of the RNA molecule above and bellow of $T_c$ and characterize the
nature of the transition that it undergoes. It is important to 
remark here
that the condition $N \ll 1$ (interpreted as meaning a large
concentration of Mg$^{++}$) is
valid for all the results presented in this section,
since this is a necessary condition for using the approximate
expression for the partition function instead of the exact one
(\ref{Zcompleta}). In this analysis, we only consider the regular
part of the partition function, disregarding the $N$-dependent,
divergent factor in (\ref{ZR}). This factor can be absorbed by
convenient renormalizations, and cancels out in any observable
quantity which involves an statistical average over the ensemble
of random configurations of the system. Therefore, all
thermodynamic quantities studied will be, therefore, termed as
`regular'.

\subsection{Free energy per base}

From (\ref{ZTA}), we determine the free energy per base,
$f=F/L$.
Using that $[L/2]/L \rightarrow 1/2$ in the large $L$ limit, we
write the regular free energy per base as:
\bea
f_R\left(L,N,T<T_c\right) &\simeq& - \frac{\kappa}{2} \ln \{ N\} T
\label{fL-Tc} \\
f_R\left(L,N,T>T_c\right) &\simeq& - \frac{\kappa}{2} \ln \{ L/e
\}T + \frac{\epsilon}{2}\ . \label{fL+Tc}
\eea
On the one hand, from (\ref{F-Tc}) we notice that $f_R$ is independent of $L$ for $T
< T_c$. This means that $f_R$ is determined by the {\it local}
(short distances) environment of a point in the chain and not by
global (large distances) properties.
This behavior is reasonable because thermal agitation is small,
which prevents coupling on a given base with another
non-neighboring one in the chain.
On the other hand,  for $T > T_c$ we see from (\ref{F+Tc})
that $f_R$ is in turn independent of $N$. This can be interpreted
by noting that for temperatures above $T_c$,
thermal agitation overcomes the folding action of the positive
ion concentration in the medium, rendering the local behavior
independent of this concentration, and, therefore, of $N$.

\subsection{Entropy}

\begin{figure}[h]
\includegraphics[width=0.53\textwidth]{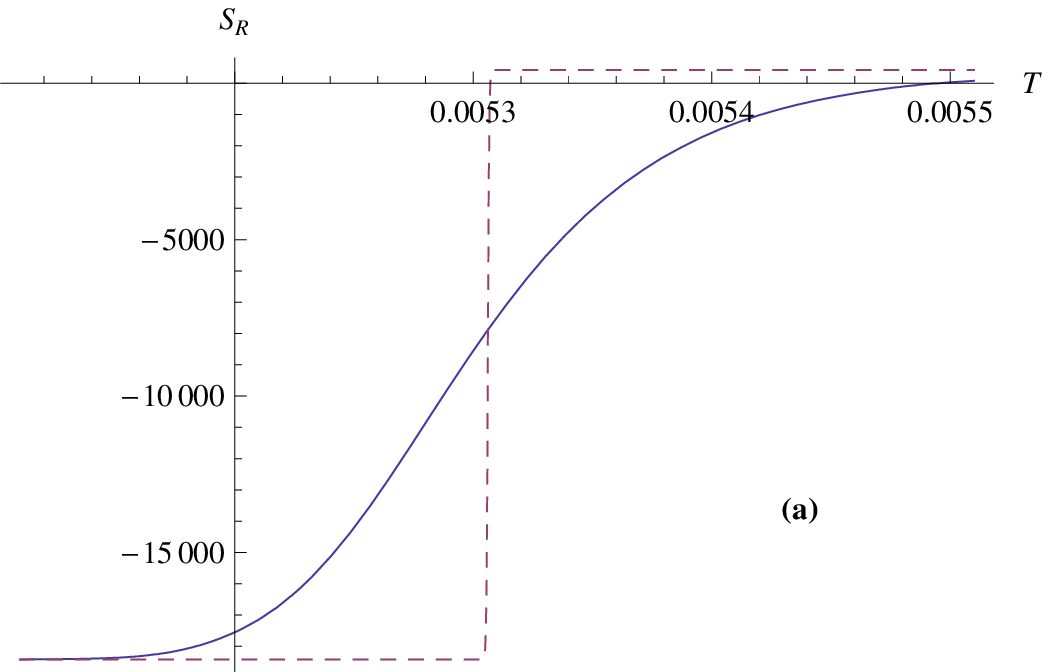}
\includegraphics[width=0.53\textwidth]{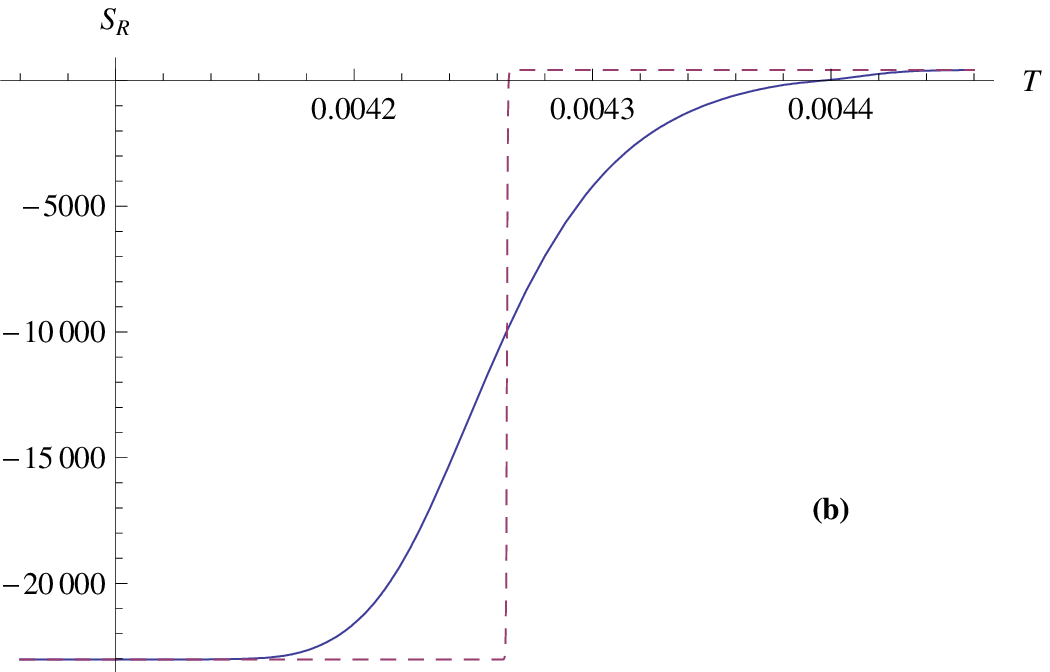}
\includegraphics[width=0.53\textwidth]{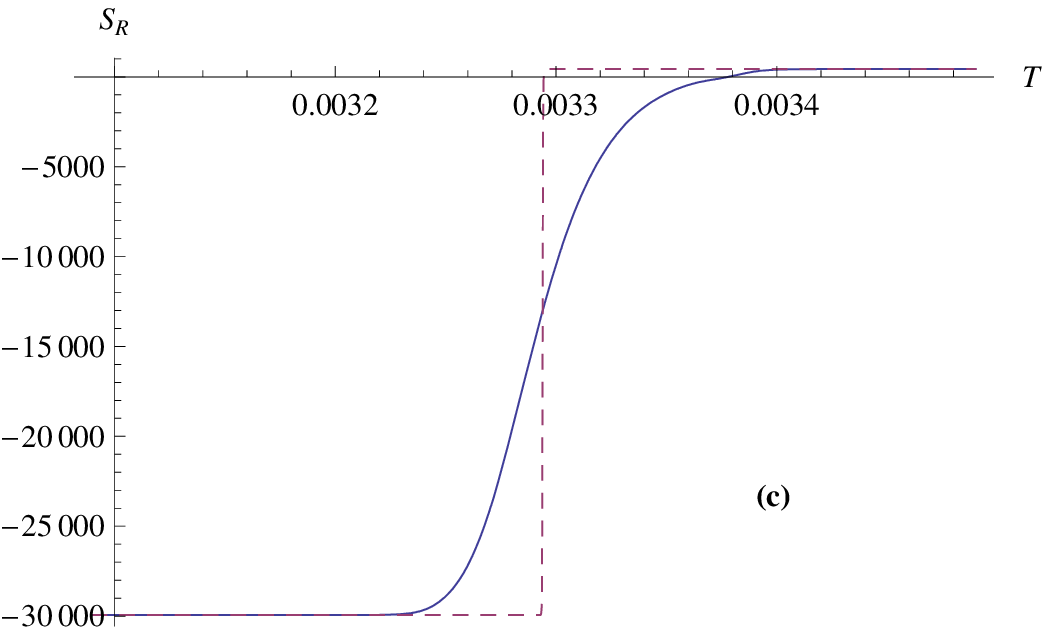}
\includegraphics[width=0.53\textwidth]{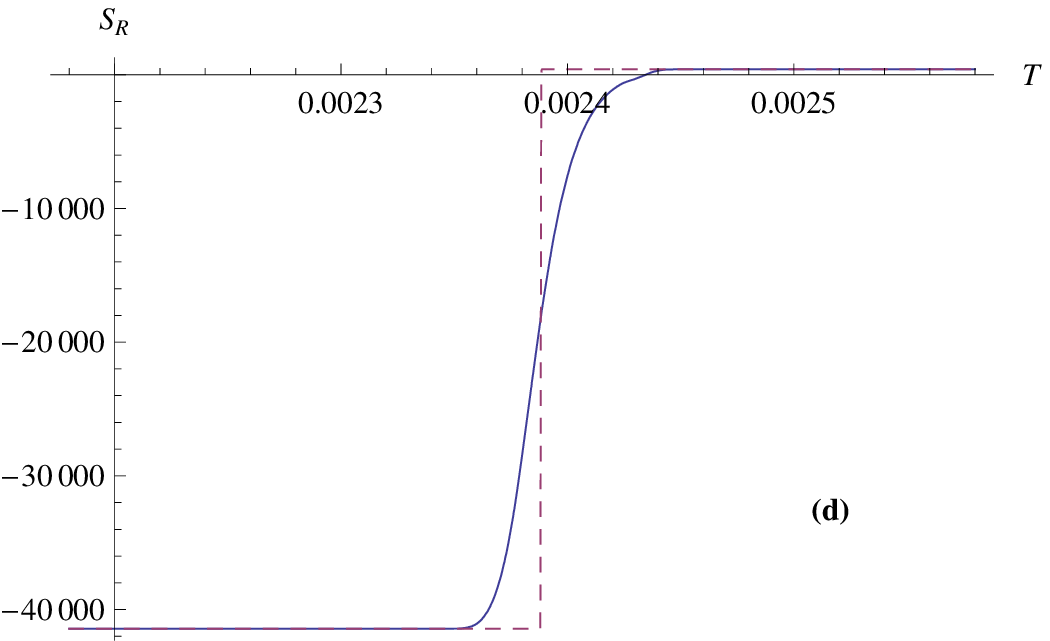}
\caption{Typical behavior of $S_R$ vs. $T$ calculated from
\textit{RZ} (continuous line) and \textit{TA} (dashed line), for
$L=200$, temperatures close to $T_c$ and different values of $N$:
(a) $N=10^{-80}$, (b) $N=10^{-100}$, (c) $N=10^{-130}$, (d)
$N=10^{-180}$. The convergence of \textit{RZ} to
\textit{TA} when $N$ decreases is displayed.} \label{Entropia}
\end{figure}

Next, we calculate the regular entropy
$S_R = -\partial F_R / \partial T$
from \textit{RZ} and \textit{TA}, and plot it in Fig. \ref{Entropia} ,
which shows that the dependence of the entropy with the
temperature resembles a step function, with a positive step in
$T=T_c$ (in particular, for the \textit{TA} case). 
This behavior
could be expected, since $S_R(T_a) \leq S_R(T_b)$ for $T_a < T_b$
and it is a consequence of the linear asymptotic behavior of the
regular free energy displayed in Fig. \ref{F2F1}. On the contrary, a
negative step in $T=T_c$ would have implied $T_c < 0$. The curves
of $S_R$ from the \textit{TA}, for $T$ below and above $T_c$, are
derived from (\ref{F-Tc}) and (\ref{F+Tc}). Moreover, it can be
seen from Fig. \ref{Entropia} that as $N$ decreases, \textit{RZ}
approaches \textit{TA} and the behavior of \textit{RZ} resembles
more closely that of a step function: 
\bea S_R
\left(L,N,T<T_c\right) &\simeq& \kappa [L/2] \ln \{ N\}\ ,
\label{S-Tc}\\
S_R \left(L,N,T>T_c\right) &\simeq& \kappa  \ln \{r_{[L/2],[L/4]}\}
\simeq \kappa [L/2] \ln \{  L/e \}\ .\label{S+Tc}
\eea
We have verified numerically that, in the limit $N \to 0$,
the plot of $S_R$ from \textit{RZ} behaves
asymptotically as step function with step in $T_c$ of magnitude
proportional to the latent heat.

\subsection{Latent heat}

From the expressions for the entropy, we calculate the regular
latent heat of the transition:
\begin{equation}
l_R \left(L,N,T=T_c\right) = T_c \Delta S_R(T_c) = T_c \{
S_R(T>T_c)-S_R(T<T_c) \}= [L/2] \epsilon\ . \label{Latent}
 \end{equation}
Eq. (\ref{Latent}) shows that the phase transition in the model we
have considered is of first order \cite{Stanley}. Note that
(\ref{Latent}) does not depend on $N$: this result could be
expected, since the latent heat is the energy released or absorbed
by the system during the transition, and $N$ is associated with
the concentration of Mg$^{++}$, and, therefore, plays the effective 
role of an external variable regardless of the way in which it has been
introduced. Eq. (\ref{Latent}) expresses that the energy exchange
between the system and the bath during the phase transition is
$[L/2] \epsilon$, since there are $[L/2]$ parings between pairs of
bases, which could be created or broken during the transition,
depending on the direction in which $T$ crosses $T_c$.

\section{Conclusions}

In this paper we have studied, using both analytical and numerical
methods, the $T-$dependent phase transition
in a simplified model of the RNA molecule, in the regime of large concentration
of positive ions in solution. This regime has similarities with the
large density phase of polymers studied by de Gennes. We have
presented an analytical expression for the critical temperature
$T_c$, which tends to zero in the thermodynamic limit. The
critical temperature separates the only configuration without
interaction between the bases (and, therefore, of genus zero)
which is dominant for $T<T_c$, from the configurations with high
energy and large genera equal to $[L/4]$, which dominate for
$T>T_c$. This transition is no to be confused with the
coil-globule transition, which appears for low concentration of
positive ions in solution. Due to the interesting dependence of
the genus with the temperature, we call this a topological phase
transition, and we have shown that the transition is of first
order. Although the model studied is very simple, we are confident
that most of the properties studied here might be robust and
extend to more realistic ones, given the importance of the
saturation property of the elementary base interaction, that this
model preserves.

{\bf Acknowledgments}

MdE thanks M. M. Reynoso for discussions.


\def\NP{{\it Nucl. Phys.\ }}
\def\PRL{{\it Phys. Rev. Lett.\ }}
\def\PL{{\it Phys. Lett.\ }}
\def\PR{{\it Phys. Rev.\ }}
\def\CMP{{\it Comm. Math. Phys.\ }}
\def\IJMP{{\it Int. J. Mod. Phys.\ }}
\def\MPL{{\it Mod. Phys. Lett.\ }}
\def\RMP{{\it Rev. Mod. Phys.\ }}
\def\AP{{\it Ann. Phys.\ }}
\def\EPL{{\it Eur. Phys. Lett.\ }}

\end{document}